\documentclass[conference]{IEEEtran}
\IEEEoverridecommandlockouts
\usepackage[utf8]{inputenc}
\usepackage{microtype}
\usepackage{graphicx}
\usepackage{booktabs}
\usepackage{color}
\usepackage{url}
\usepackage{cite}

\title{A General Model for Deepfake Speech Detection:\\
Diverse Bonafide Resources or Diverse AI-Based Generators}

\author{
Lam~Pham$^{1*}$,
Khoi~Vu$^{2*}$,
Dat~Tran$^{2*}$,
David~Fischinger$^{1}$,\\
Alexander~Schindler$^{1}$,
Martin~Boyer$^{1}$,
Ian~McLoughlin$^{3}$%
\thanks{L. Pham, D. Fischinger, A. Schindler, and M. Boyer are with Austrian Institute of Technology, Austria.}%

\thanks{K. Vu and D. Tran are with FPT University, Vietnam.}%

\thanks{I. McLoughlin is with Singapore Institute of Technology, Singapore.}%

\thanks{(*) Main and equal contribution into the paper.}
}

\begin{document}
\maketitle

\begin{abstract}
In this paper, we analyze two main factors of Bonafide Resource (BR) or AI-based Generator (AG) which affect the performance and the generality of a Deepfake Speech Detection (DSD) model. 
To this end, we first propose a deep-learning based model, referred to as the baseline. Then, we conducted experiments on the baseline by which we indicate how Bonafide Resource (BR) and AI-based Generator (AG) factors affect the threshold score used to detect fake or bonafide input audio in the inference process. 
Given the experimental results, a dataset, which re-uses public Deepfake Speech Detection (DSD) datasets and shows a balance between Bonafide Resource (BR) or AI-based Generator (AG), is proposed.
We then train various deep-learning based models on the proposed dataset and conduct cross-dataset evaluation on different benchmark datasets.
The cross-dataset evaluation results prove that the balance of Bonafide Resources (BR) and AI-based Generators (AG) is the key factor to train and achieve a general Deepfake Speech Detection (DSD) model.
\end{abstract}

\section{INTRODUCTION}
Over the past few years, generative AI presents an impressive development due to remarkable advancements in deep-learning techniques.
Regarding applying generative AI in audio domain, especially focusing on human speech, a wide range of high-quality applications such as media production, AI-based music composer, audio-based chatbot, etc. have been introduced and used popularly.
However, this also poses potential risks when generated human speech by AI-based generators is used for criminal purposes.
To address this concern, the audio forensic task of Deepfake Speech Detection (DSD) has been raised and drawn much attention from the audio research community.
Indeed, various public datasets from scientific papers and challenge competitions have been proposed for the DSD task recently (i.e., The survey paper~\cite{pham_01} indicates that four DSD datasets each year have been published from 2021).
Given the public datasets, a wide range of deep-learning based models have been proposed to detect the deepfake audio, referred to as DSD models.
The state-of-the-art DSD models~\cite{sota_01_xlsr, sota_02, sota_03} have leveraged pre-trained deep neural networks which were trained on large-scale datasets of human speech for the up-stream task of Speech-to-Text.
Then, these pre-trained models are fine-tuned on the down-stream task of DSD.
In particular, authors in~\cite{sota_01_xlsr} achieved the best performance by fintuning XLSR model, a variant of Wave2Vec2~\cite{wav2vec20} released by Meta.
Meanwhile, authors in~\cite{sota_02} leveraged and finetuned WavLM~\cite{wavlm} released by Microsoft.
The authors in~\cite{sota_03} made effort to benchmark a wide range of popular pre-trained models proposed for audio tasks.


However, the state-of-the-art DSD models present two main concerns.
First, these DSD models use Equal Error Rate (EER) with free threshold score as the main metric to compare the performance among proposed models.
The EER scores are reported and treated as individual values as they are computed from different testing datasets rather than on all testing datasets.
For each EER value obtained from a specific testing dataset, there is a corresponding threshold score which is used to detect fake or bonafile audio.
These threshold scores may be significantly different from different testing datasets.
However, only one threshold score must be selected to detect fake or bonafile audio from an unseen data in the inference process.
Therefore, if the threshold scores are much different from different testing datasets, it leads the overfitting issue and the DSD model is not general.
In other words, proposed high-performance DSD models with low EER values evaluated on individual testing datasets may be not really reliable if the corresponding threshold scores are not presented, compared, and analyzed.
However, the state-of-the-art DSD systems have not reported or analyzed threshold scores across testing datasets.
\begin{figure}[t]
    \centering
    \scalebox{0.5}{
    \includegraphics[width=\textwidth]{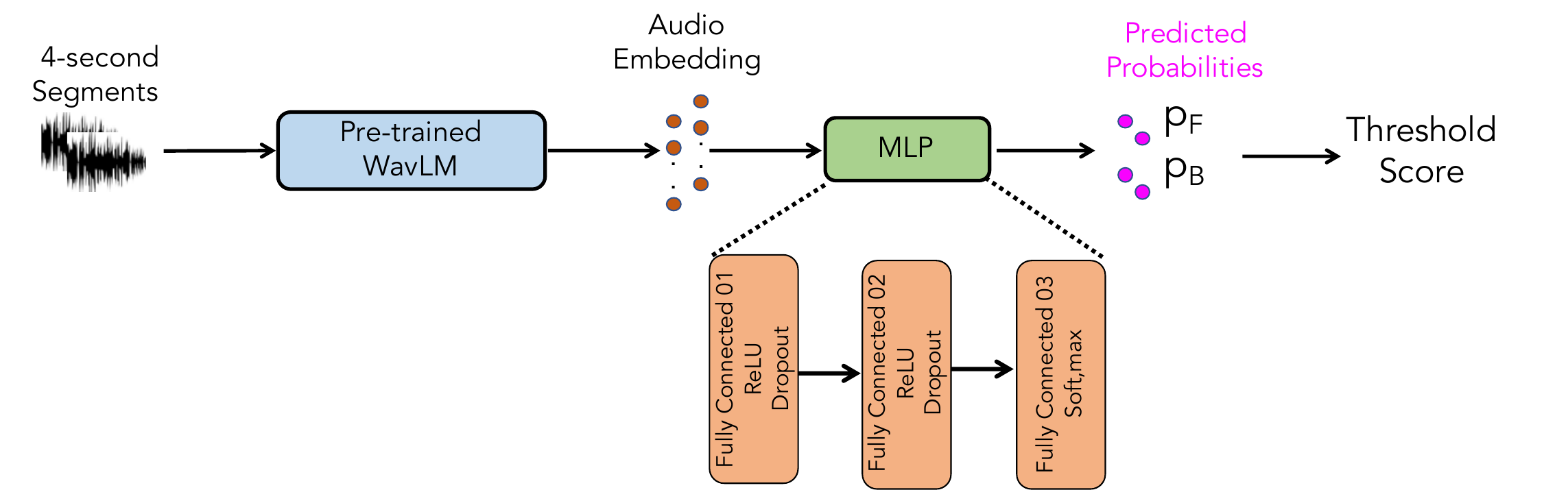}
    }
    \caption{The proposed baseline architecture}
    \label{fig:baseline}    
\end{figure}

Second, during training a DSD model, Softmax-based layer is normally used to separate fake and bonafide audio into two distributions with the boundary probability of 0.5. However, the unbalance issues of Bonafide Resource (BR) or AI-based Generators (AG) may lead the fake or bonafide distributions to mainly locate at the boundary (e.g., around the probability of 0.5). Therefore, increasing or decreasing the probability boundary (e.g., upper or lower than 0.5) helps to find the optimized boundary probability from which the EER and the corresponding threshold score are computed. 
However, none of research has analyzed the role of Bonafide Resource (BR) or AI-based Generators (AG) which may significantly affect the final probability boundary.   
 
By addressing these concerns, we make the following main contributions in this paper:
\begin{itemize}
    \item We conduct extensive experiments to indicate the roles of Bonafide Resource (BR) and AI-based Generator (AG) which affect a DSD model performance, the EER metric and the corresponding threshold score.
    \item Given the experimental results, we propose a dataset for DSD task which re-uses public DSD datasets and shows a balance between Bonafide Resource (BR) and AI-based Generator (AG).
    \item With the proposed dataset, we train various deep-learning based model for DSD task. The proposed models proves general when we conduct cross-dataset evaluation on different benchmark datasets and using a fixed threshold. 
\end{itemize}

\section{Evaluation of Bonafide Resource and AI-Based Generator Factors}
\subsection{Proposed Baseline Model}
\label{sec11}

\begin{table*}[t]
    \caption{Public and Benchmark Datasets Proposed for Deepfake Speech Detection (DSD) in English} 
        	\vspace{-0.2cm}
    \centering
    \scalebox{0.7}{
    \begin{tabular}{|l|c|c|c|c|c|} 
       \hline  
       \textbf{Datasets}   &\textbf{Years}  &\textbf{Speakers} &\textbf{Utt. No.} &\textbf{Fake speech}   &\textbf{Real Speech} \\  
                           &                &(Male/Female)     &(Bonafide/Fake)       &\textbf{Generators}    &\textbf{Resources}                          \\
       \hline     
       \hline     
       ASVspoof 2015~\cite{ASV15}(audio)            
       &2015            
       &45/61           
       &16,651/246,500   
	    &10 (7 VC, 3 TTS)                         
       &Speaker Volunteers          \\
       \hline
       
       FoR~\cite{for_ds}(audio)                     
       &2019            
       &140                
       &-/195541         
       &7 TTS                            
       &Kaggle~\cite{for_source}   \\
       \hline
       
       ASVspoof 2019 (LA task)~\cite{ASV19}(audio)  
       &2019            
       &46/61           
       &12,483/108,978          
	    &19 (8VC, 11 TTS)                          
       &Speaker Volunteers      \\     
       \hline
       
       DFDC~\cite{dfdc_ds}(video)           
       &2020            
       &3426              
       &128,154/104,500
       &1 TTS                           
       &Speaker Volunteers      \\           
       \hline
       
       ASVspoof 2021 (LA task)~\cite{ASV21}(audio)  &2021        
       &21/27           
       &18,452/163,114                
       &13 TTS/VC                          
       &Speaker Volunteers    \\      
       \hline
       
       \textbf{ASVspoof 2021 (DF task)~\cite{ASV21}(audio)}  &2021        
       &21/27           
       &\textbf{22,617/589,212}                 
       &\textbf{100+ TTS/VC}                        
       &\textbf{Speaker Volunteers}   \\       
       \hline
  
       WaveFake~\cite{wake_ds}(audio)               
       &2021            
       &0/2             
       &-/117,985         
       &6 TTS                          
       &LJSPEECH~\cite{ljspeech_ds},   \\ 
       & &  & & &JSUT~\cite{jsut_ds}  \\
       \hline
    
       FakeAVCeleb~\cite{khalid2021fakeavceleb}(video) &2022  &250/250         &570/25,000            &2 TTS                           &Vox-Celeb2~\cite{voxceleb2} \\ 
       \hline

       In-the-Wild~\cite{intwi_ds}(video)           
       &2022              
       &58                
       &19963/11816              
       &n/a                          
       &Self-collected          \\

       \hline
       
       Voc.v~\cite{VoC_ds}(audio)                   
       &2023            
       &46/61           
       &14,250/41,280             
       &5 TTS                           
       &ASVspoof 2019    \\       
       \hline

       
       LibriSeVoc~\cite{f_ds_01}(audio)             
       &2023            
       &n/a                 
       &13,201/79,206     
       &6 TTS/VC                           
       &Librispeech~\cite{libreispeech}    \\                
       \hline

       \textbf{AV-Deepfake1M~\cite{cai2023av, 1mdeepfake_ch}(video)}   
       &2023 
       &2,068              
       &\textbf{286,721/860,039}  
       &\textbf{2 TTS}                           
       &\textbf{Voxceleb2~\cite{voxceleb2}}      \\               
       \hline
     
        MLAAD~\cite{muller2024mlaad}(audio)         
        &2024            
        &n/a                  
        &-/76,000             
        &54 TTS                          
       &Librispeech~\cite{libreispeech}         \\           
       \hline

    \textbf{ASVspoof5~\cite{ASV24}(audio)}     
    &2024            
    &n/a                  
    &\textbf{188,819/815,262}             
    &\textbf{28 (15 TTS, 6 VC, 7 AT)}        
    &\textbf{Librispeech~\cite{libreispeech}}       \\             
    \hline

    \end{tabular}
    }
    \vspace{-0.3cm}
    \label{table:dataset} 
\end{table*}

To evaluate the role of Bonafide Resource (BR) and AI-Based Generator (AG) to affect the DSD model performance, we first propose a deep-learning based model which is referred to as the baseline in this paper.
Inspired by leveraging pre-trained models~\cite{sota_02, sota_03, sota_01_xlsr} from the state-of-the-art, we also use the pre-trained WavLM~\cite{wavlm} model to construct the baseline.
In particular, the baseline is shown in Fig.~\ref{fig:baseline}.
First, 4-second audio segments are fed into the pre-trained WavLM~\cite{wavlm} model which was trained on large-scale datasets for the task of Speech-to-Text.  
By leveraging the pre-trained WavLM model~\cite{wavlm}, audio embeddings (i.e., The audio embedding is a vector-based presentation) extracted from this model present general acoustic features of human speech. 
Then audio embeddings are explored by a Multilayer Perceptron (MLP) which is performed by three Dense layers. 
During training the baseline, the pre-trained WavLM model is frozen.
In other words, only trainable parameters of the MLP component is updated during the training process to adapt the DSD task.     

\subsection{The Statistics of Public and Benchmark DSD Datasets}

\begin{figure}[t]
    \centering
    \scalebox{0.4}{
    \includegraphics[width=\textwidth]{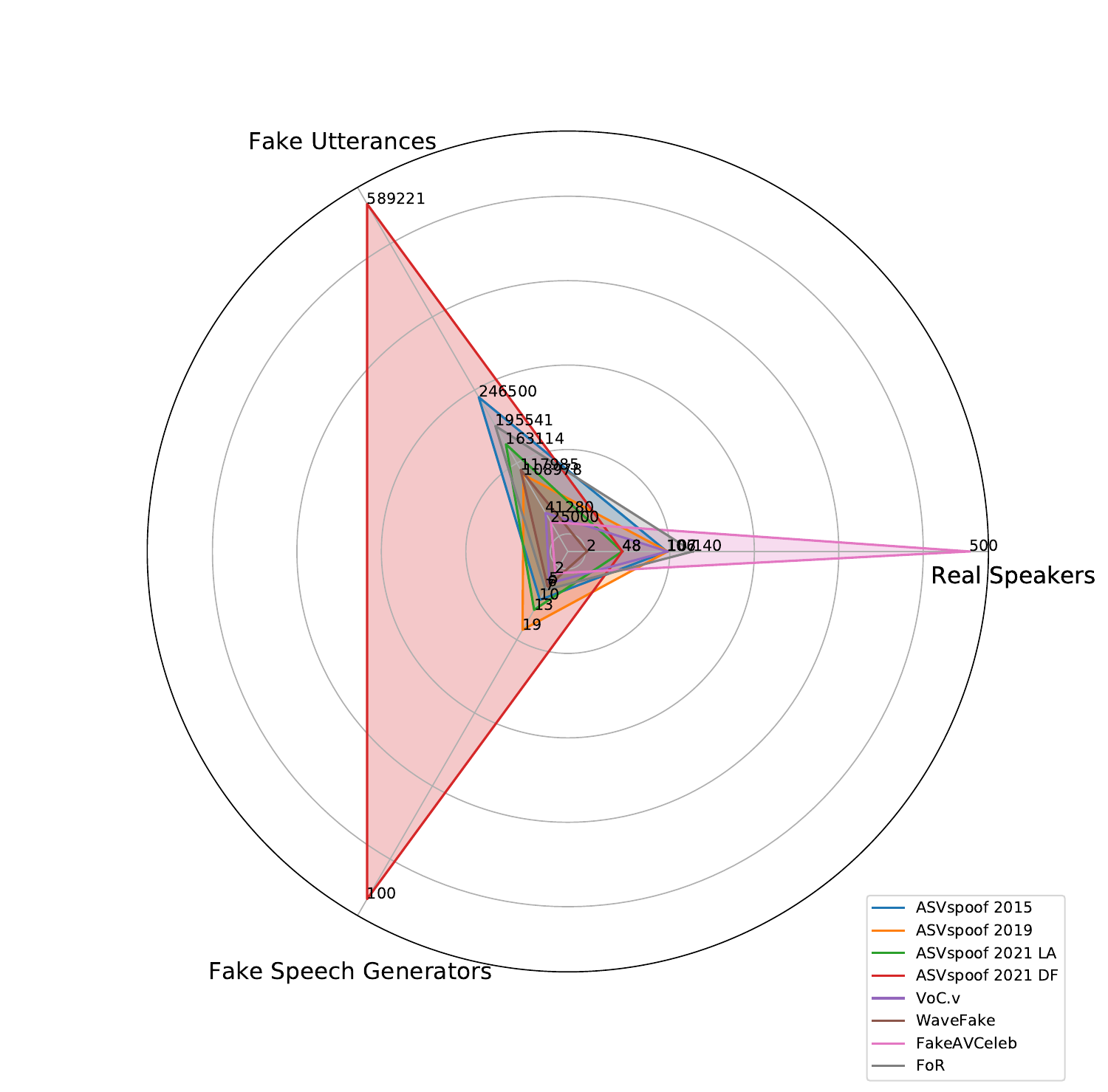}
    }
    \caption{The statistics of DSD datasets in English}
    \label{fig:dataset}    
\end{figure}

We now analyze the statistics of public and benchmark datasets proposed for the DSD task (i.e., Only English datasets are analyzed).
To this end, we present the public datasets for the DSD task in Table~\ref{table:dataset} which were also summarized in our previous work~\cite{pham_01}.
The Table~\ref{table:dataset} indicates an unbalanced issue between Bonafide Resource (BR) and AI-Based Generator (AG).
Indeed, ASVspoof 2021 (DF task)~\cite{ASV21} presents the largest number of AG (e.g., more than 100 Text-to-Speech (TTS) and Voice Conversion (VC)) but is limited by the number of speakers (e.g., 21 males and 27 females). 
In contrast, some datasets such as ASVspoof5~\cite{ASV24}, WaveFake~\cite{wake_ds}, LibriSe Voc~\cite{f_ds_01} leverage the  Librispeech~\cite{libreispeech}, a large-scale of human speech dataset, as the bonafide resource. 
This is effective to construct a general human speech dataset with diverse speech characteristics such as gender, accent, loudness, speech flow, etc. that reflects real-world conversations.
However, these datasets present a limitation of AG and unbalanced issue between TTS and VC systems.
Similarly, AV-Deepfake1M~\cite{cai2023av} presents a large number of fake utterances but fake utterances are generated from only 2 TTS systems. 
The Fig.~\ref{fig:dataset} again indicates an unbalance issue among Bonafide Resource (BR), AI-Based Generators (AG), and the fake utterance number regarding proposed DSD datasets without using the Librispeech as the bonafide resource.
\begin{table}[t]
    \caption{Proposed datasets to balance AI-Based Generator and Bonafide Resource factors} 
        	\vspace{-0.2cm}
    \centering
    \scalebox{0.8}{
    \begin{tabular}{|c|c|c|} 
       \hline  
       \textbf{Datasets}  &\textbf{Bonafide Utterances} &\textbf{Fake Utterances } \\
       \hline     
       \hline   
        AG dataset &53,552 &861,304 \\                  
        \hline 
       BR dataset &50,131 &273,176 \\                  
       \hline
        BR-AG dataset &103,683 &1,134,408 \\                  
       \hline         

    \end{tabular}
    }
    \vspace{-0.3cm}
    \label{table:proposed_data} 
\end{table}
\subsection{Experimental Settings} 
\label{first-exp}

Given the DSD dataset concerns above, we now setup an experiment to evaluate the roles of Bonafide Resource (BR) and AI-Based Generators (AG).
First, ASVspoof 2019 (LA), ASVspoof 2021 (LA), and ASVspoof 2021 (DF) are grouped into one dataset with totally 53,552/861,304 of bonafide/fake utterances.
We gather these datasets as they use bonafide audio from a limited speaker volunteers but apply diverse AI-Based Generators (e.g., more than 100 generators).
This data is referred to as the AG dataset
Meanwhile, ASVspoof 2024~\cite{ASV24} train and development subsets are also grouped into one dataset with totally 50,131/273,176 of bonafide/fake utterances. 
As these datasets use diverse bonafide audio from Libreispeech~\cite{libreispeech}, these are group together to present a diverse bonafide dataset, referred to as the BR dataset.
\begin{figure}[t]
       \vspace{-0.3cm}
    \centering
    \scalebox{0.4}{
    \includegraphics[width=\textwidth]{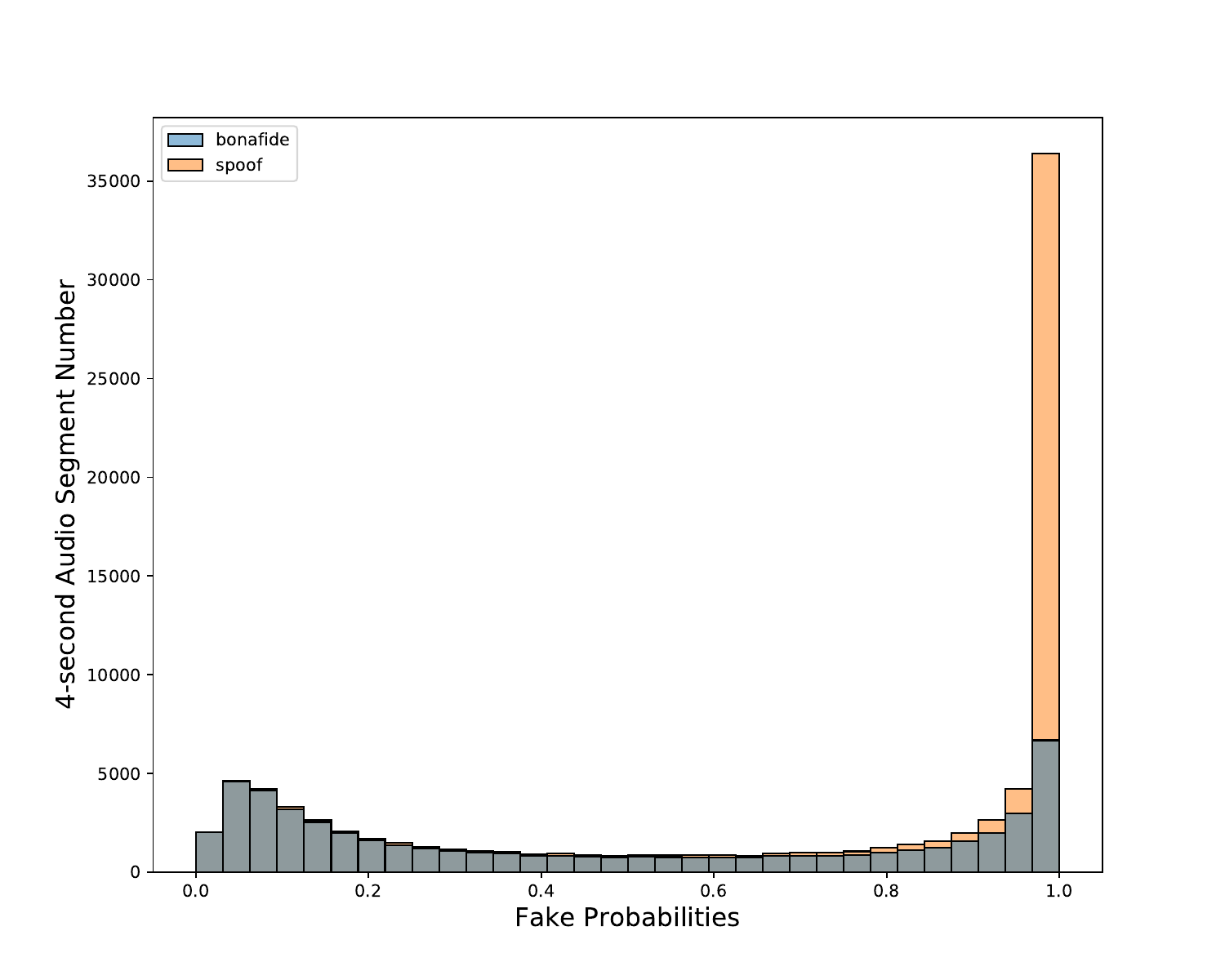}
    }
       \vspace{-0.3cm}
    \caption{Distribution of fake probabilities from In-The-Wild dataset using AG dataset to train the baseline model}
    \label{fig:dis01}    
\end{figure}

\begin{figure}[t]
       \vspace{-0.5cm}
    \centering
    \scalebox{0.4}{
    \includegraphics[width=\textwidth]{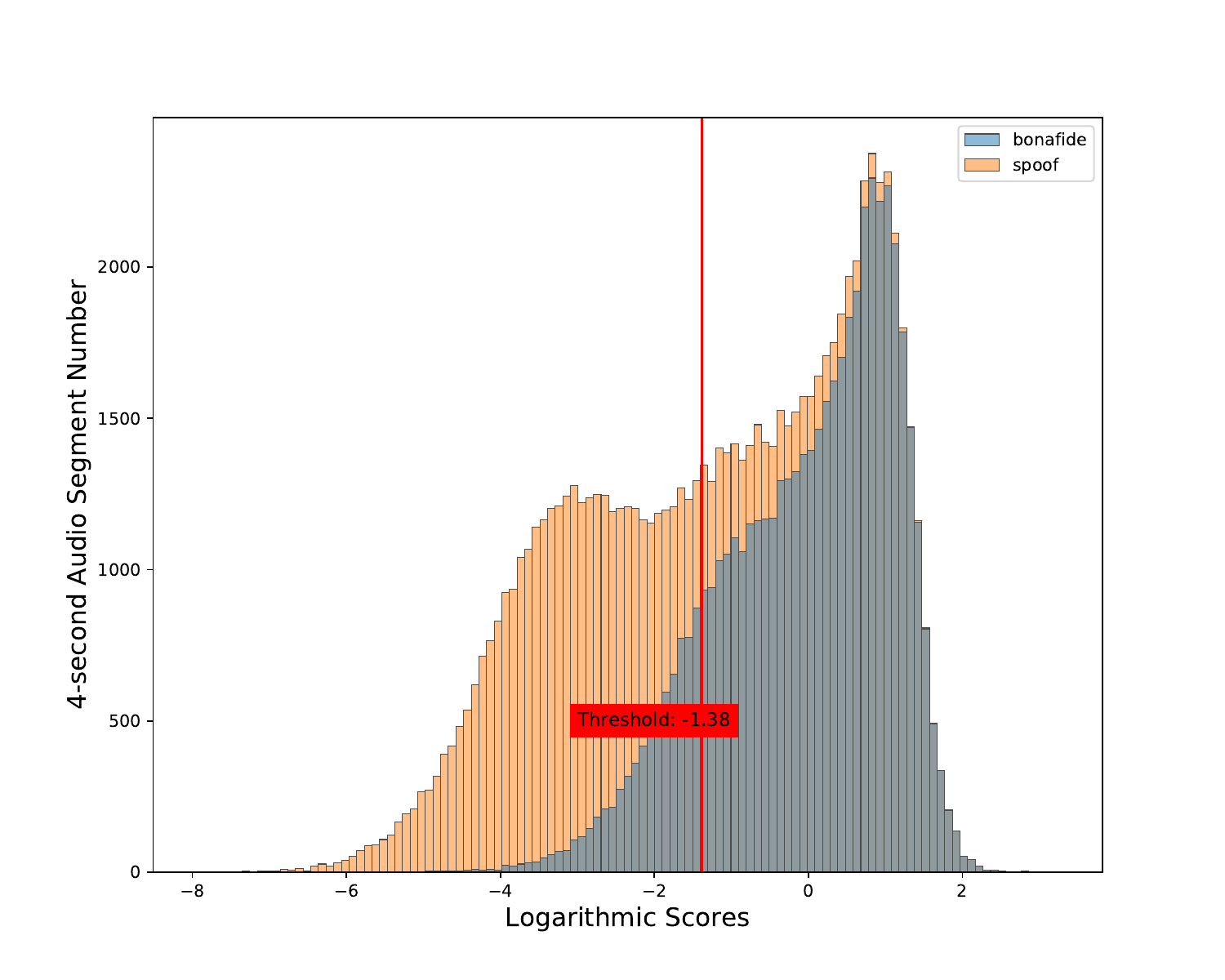}
    }
       \vspace{-0.3cm}
    \caption{Distribution of Logarit scores from In-The-Wild dataset using AG dataset to train the baseline model}
    \label{fig:dis03}    
\end{figure}
Given the BR dataset and the AG dataset as shown in Table~\ref{table:proposed_data}, we setup two test cases.
In the first test case, we evaluate the role of AG.
We use the AG dataset to train the baseline model. After the training process, we test the baseline model on the In-The-Wild~\cite{intwi_ds} dataset.
In the second test case, we evaluate the role of BR.
We use the BR-dataset to train the baseline model. Then, we test the baseline model on the In-The-Wild~\cite{intwi_ds} dataset.
We select In-The-Wild~\cite{intwi_ds} for testing in both test cases as this dataset has been widely used for cross-dataset evaluation in the state-of-the-art papers and has not presented the AG or BR information. This dataset also presents various noise background, different audio quality which reflect real scenarios, and balance between 52,605 4-second bonafide segments and 35483 4-second fake segments. 
\begin{figure}[t]
       \vspace{-0.3cm}
    \centering
    \scalebox{0.4}{
    \includegraphics[width=\textwidth]{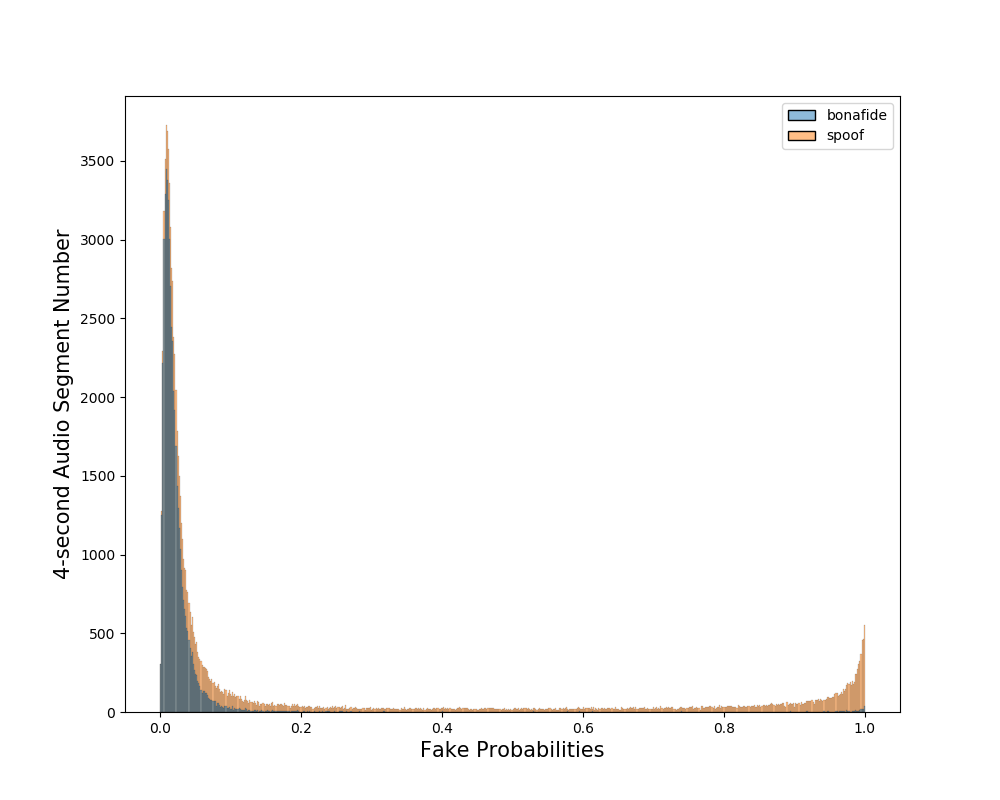}
    }
       \vspace{-0.3cm}
    \caption{Distribution of fake probabilities from In-The-Wild dataset using BR dataset to train the baseline model}
    \label{fig:dis02}    
\end{figure}

\begin{figure}[t]
       \vspace{-0.5cm}
    \centering
    \scalebox{0.4}{
    \includegraphics[width=\textwidth]{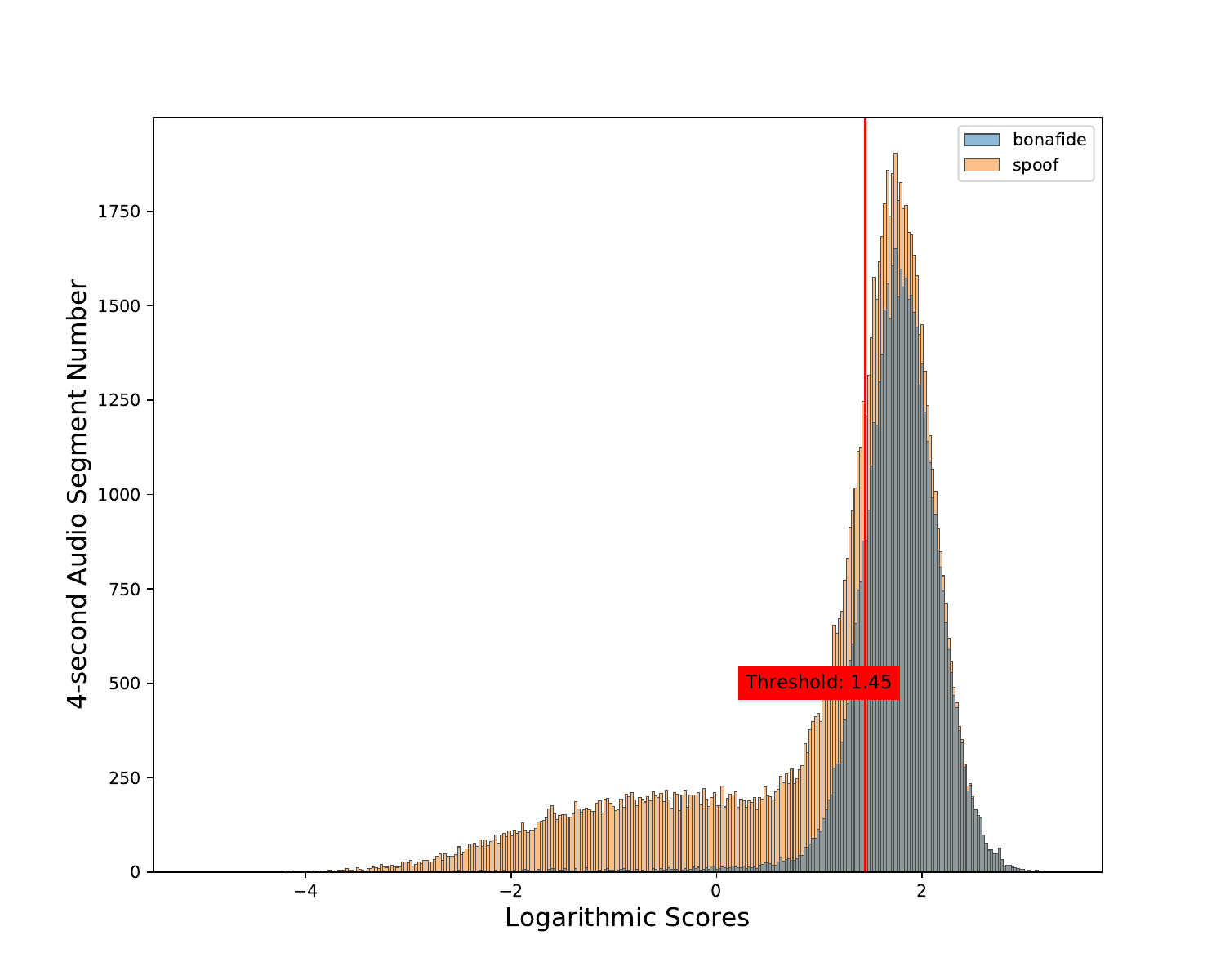}
    }
           \vspace{-0.3cm}

    \caption{Distribution of Logarit scores from In-The-Wild dataset using BR dataset to train the baseline model}
    \label{fig:dis04}    
\end{figure}
The baseline model is constructed with Pytorch framework and is trained on GPU GeForce 12 GB.
We uses Adam~\cite{Adam} algorithm for optimization. 
The training process is stop after 30 epoch and uses a fixed learning rate of 1E-5.
Notably, this experiment is for 4-second audio segment input. 
In other words, original audio utterances are split into 4-second audio segments before feeding into the model. 

Regarding evaluating metrics, we present F1 score, Accuracy, AuC, and EER. 
The EER is accompanied with the threshold score. 
We use the probability of 0.5 to compute Accuracy, F1 Score and AuC (i.e., if the fake probability is larger than 0.5, the input audio is referred to as fake).
Meanwhile, we follow ASVspoof challenge to compute EER and the corresponding threshold basing on the Logarithmic scores. 
Particularly, the Logarithmic score is computed from fake and bonafide probabilities in the following equation:
\begin{equation}
    \label{eq:score}
    Log\_Score = \log_{10}^{(P_{B}/P_{F})},
\end{equation}
where $P_B$ and $P_F$ are the bonafide and fake probabilities, respectively.

\begin{figure}[t]
    \centering
    \scalebox{0.4}{
    \includegraphics[width=\textwidth]{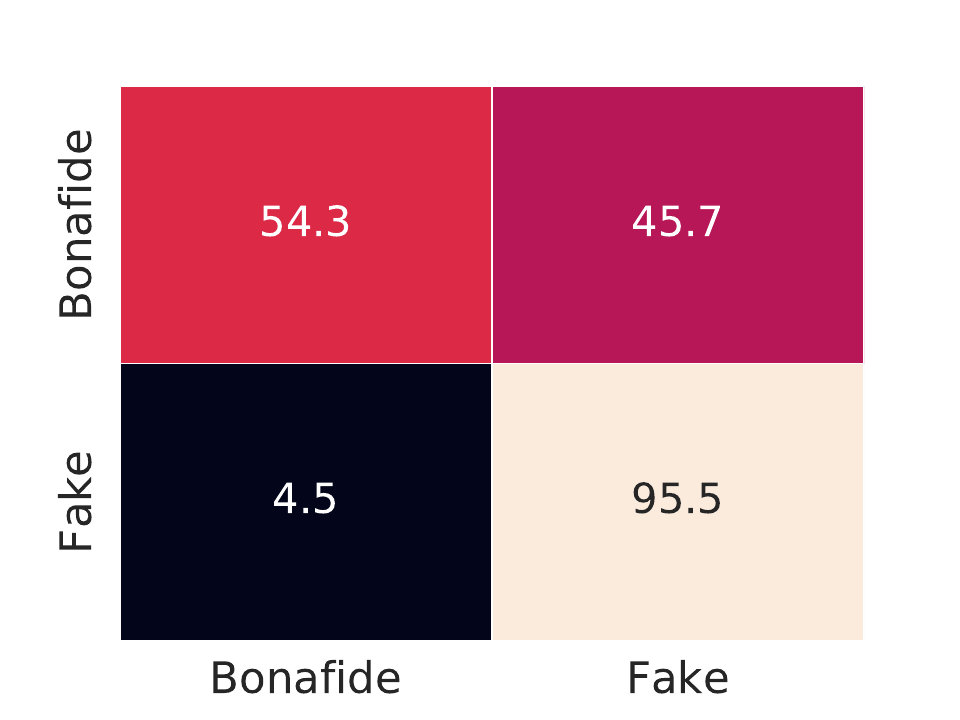}
    }
    \caption{Confusion matrix (\%) for In-The-Wild dataset using the AG dataset to train the baseline model (The boundary probability of 0.5)}
    \label{fig:con01}    
\end{figure}

\subsection{Experimental Results and Discussion}
\label{sec13}

As Fig.~\ref{fig:dis01} and Fig.~\ref{fig:dis03} show, this indicates that the distribution of fake probabilities (e.g., the output of Softmax layer) from fake audio locates near one when we train the baseline on AG-dataset.
Meanwhile, the distribution of fake probabilities from bonafide audio spreads from zero to near one.
This leads the threshold of -1.38 when EER of 0.85 is computed.
In other words, the fake probability is significant larger than the bonafide probability (e.g., fake probability is near one) to decide a fake audio input if we use the threshold score of -1.38.

In contrast, when we train the baseline on the BR-dataset, the distribution of fake probabilities from fake audio spread from zero to one while the fake probabilities from bonafide audio locate near zero. 
This leads the threshold value of 1.45 when the EER of 0.82 is computed.
In other words, the fake probability is significant smaller than bonafide probability (e.g., the fake probability is near zero) to decide a fake audio input if we use the threshold score of 1.45 as the boundary.
\begin{figure}[t]
    \centering
    \scalebox{0.4}{
    \includegraphics[width=\textwidth]{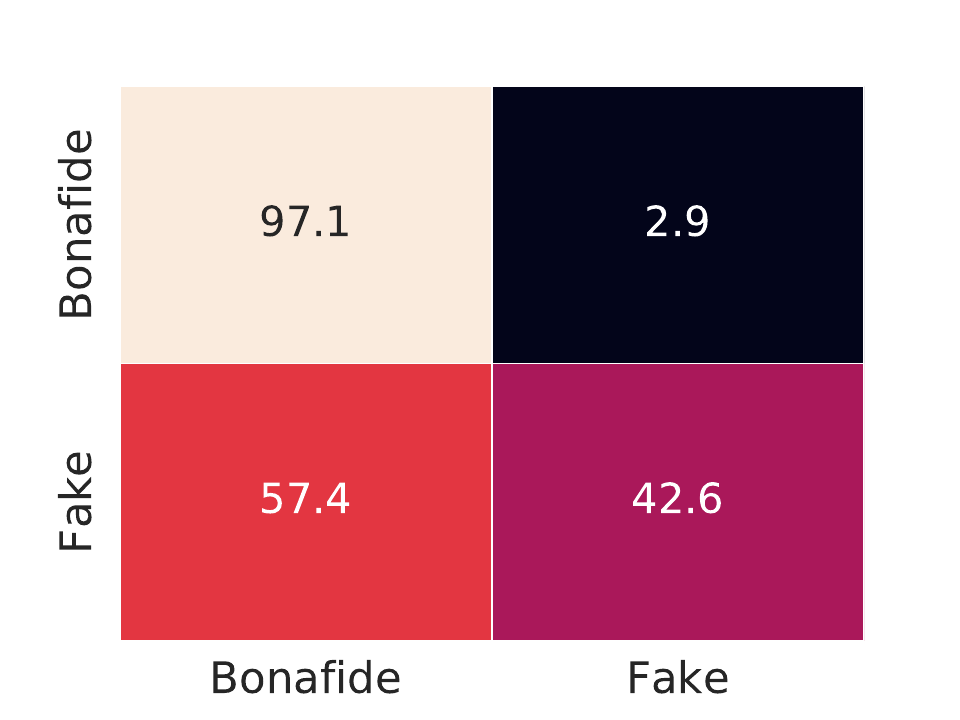}
    }
    \caption{Confusion matrix (\%) for In-The-Wild dataset using the BR dataset to train the baseline model (The boundary probability of 0.5)}
    \label{fig:con02}    
\end{figure}

As a result, we conclude that if we train a model on either AG-dataset or BR-dataset and then use the threshold score accompanied with EER metric for the inference process, this leads the overfitting issue on the training dataset and may obtain significant misclassification on unseen data.
Indeed, instead of using the threshold score from the EER metric, we use the probability of 0.5 (e.g., The output of Softmax layer) as the probability boundary to decide fake or bonafide audio input, the confusion matrix shown in Fig.~\ref{fig:con01} and Fig.~\ref{fig:con02} indicate that the misclassification occurs on bonafide audio if training on AG dataset and on fake audio if training on BR dataset.

\begin{figure*}[t]
    \centering
    \scalebox{0.9}{
    \includegraphics[width=\textwidth]{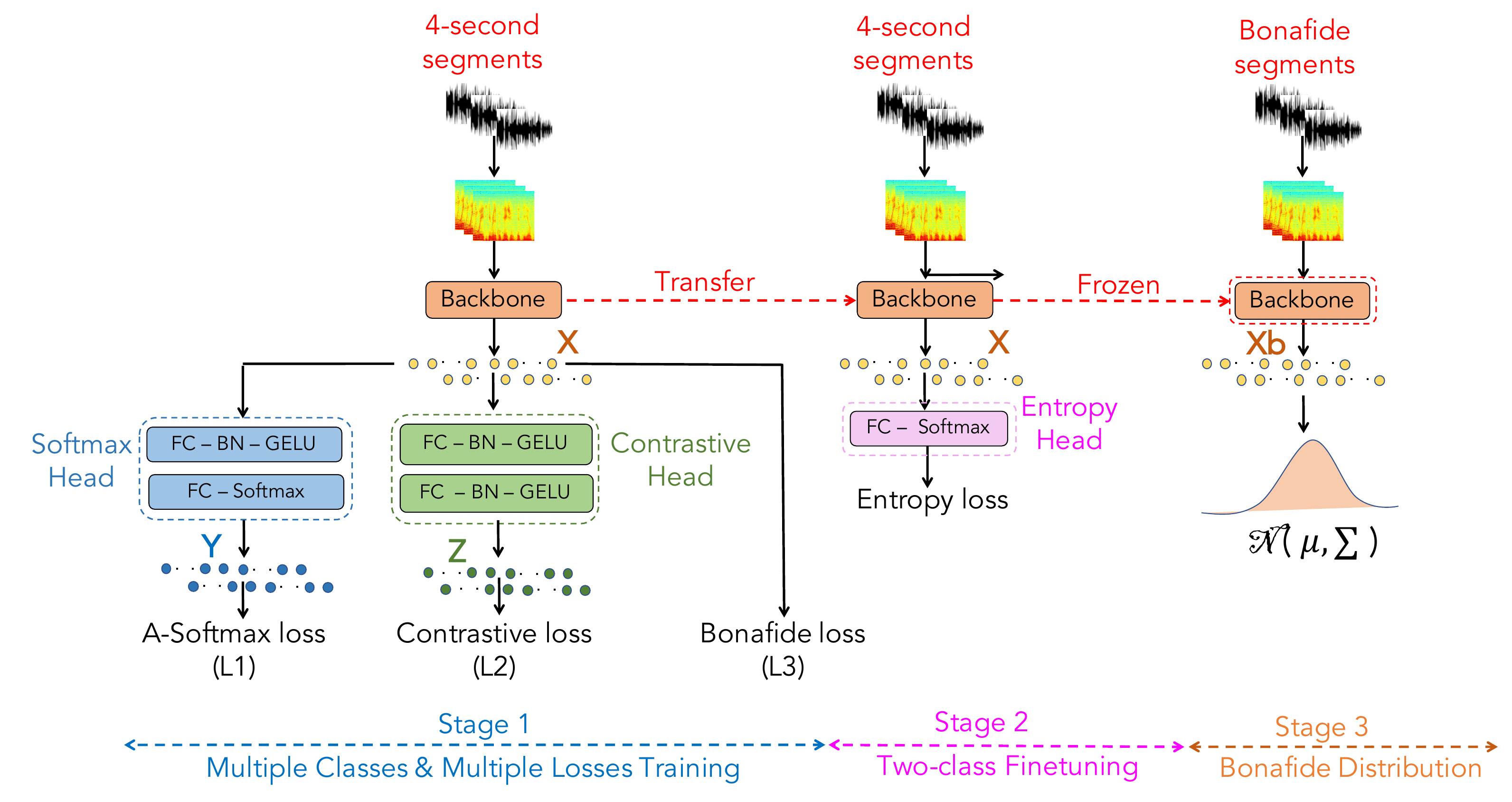}
    }
    \caption{Our proposed model with three-stage training strategy for DSD task}
    \label{fig:model}    
\end{figure*}

\section{Proposed DSD Dataset, DSD model, and Cross-Test Evaluation}

\subsection{Proposed Dataset for DSD task} 
Experimental results in the Section~\ref{sec13} indicate that the unbalance issue between Bonafide Resource (BR) and AI-Based Generators (AG) leads significantly different threshold scores. Consequently, a DSD model, which was trained on unbalance datasets of BR and AG, may be not general for unseen data in the inference process if using the threshold score from EER metric.
In other words, individual EER scores and corresponding threshold scores evaluated on independent testing datasets may not exactly reflect the generality of a DSD model. 
%
To address these concerns, we proposed a DSD dataset which comprises both BR-dataset and AG-dataset.
The dataset is referred to as the BR-AG dataset.
As Table~\ref{table:proposed_data} shows, the BR-AG dataset comprises 103,683 bonafide utterances and 1,134,480 fake utterances. 
Regarding fake utterances, there are 462,354 utterances generated from Text-to-Speech (TTS) systems and 672,126 utterances generated from Voice Conversion (VC) systems.

\subsection{Proposed a DSD model}
\label{{pro_model}}

Given the proposed BR-AG dataset, we construct a DSD model which improves the proposed baseline architecture in Section~\ref{sec11} and re-uses the training strategy in our previous published work~\cite{novel_train}.
As Fig.~\ref{fig:model} shows, training the proposed DSD models comprises three main stages.
In the first stage, the pre-trained model is used as the backbone to extract audio embeddings. 
Then, the audio embeddings are explored by three heads, each of which is applied by one loss function.
The first loss function is A-Softmax loss which is used to classify audio embeddings into three categories: bonafide, TTS (fake audio generated from TTS systems), and VC (fake audio generated from VC systems).
The second loss function is Contrastive Loss which is used to separate bonafide audio embeddings with fake audio embeddings.
The final loss function, Central Loss, is used to condense the bonafide audio embeddings.
The combination of theses three loss function is effective to separate distribution of fake and bonafide audio embeddings far away.
In the second stage, three heads in the first stage are removed. The backbone is connected to only one head with Binary-Cross-Entropy loss to learn two categories of fake and bonafide. 
In the final stage, bonafide utterances are fed into the backbone to extract bonafide audio embeddings. These audio embeddings are used to compute the bonafide distribution using Mahalanobis distribution. 
In the inference process, the unseen audio is first fed into the backbone to extract the audio embedding. The the audio embedding is compared with the bonafide distribution to decide whether the unseen audio is fake or bonafide.

\begin{table*}[t]
    \caption{Evaluation of the BR and AG factors on In-The-Wild dataset (Using Probability Boundary of 0.5)} 
        	\vspace{-0.2cm}
    \centering
    \scalebox{0.9}{
    \begin{tabular}{|l|c|c|c|c|c|} 
       \hline  
       \textbf{Model} & \textbf{Train Data}  &\textbf{Test Data} &\textbf{Acc.} &\textbf{F1}  &\textbf{AuC}\\
       \hline     
       \hline   
       WavLM-Large frozen + MLP (baseline) & AG dataset &In-The-Wild &0.71 &0.72 &0.74 \\                  
\hline  
       WavLM-Large frozen + MLP (baseline) & BR dataset &In-The-Wild &0.75 &0.58 &0.69 \\                  
\hline
\hline
       WavLM-Large frozen + MLP (baseline) & BR-AG dataset &In-The-Wild &0.80 &0.72 &0.82 \\                  
       \hline
       WavLM-Large finetune + MLP & BR-AG dataset &In-The-Wild &\textbf{0.87} &\textbf{0.79} &\textbf{0.83} \\
       \hline
       Whisper-base finetune + MLP & BR-AG dataset &In-The-Wild &0.73 &0.60 &0.65 \\
       \hline
       Wav2Vec2-XLSR-53 finetune + MLP & BR-AG dataset &In-The-Wild &0.79 &0.64 &0.73 \\
       \hline         

    \end{tabular}
    }
    \vspace{-0.3cm}
    \label{table:res01} 
\end{table*}

\begin{table}[t]
    \caption{Cross-Dataset Evaluation of the best model (WavLM-Large finetune + MLP) using the boundary probability of 0.5} 
        	\vspace{-0.2cm}
    \centering
    \scalebox{0.9}{
    \begin{tabular}{|c|c|c|c|} 
       \hline  
       \textbf{Test Dataset}   &\textbf{Accuracy}  &\textbf{F1 Score} &\textbf{AuC}\\
       \hline     
       \hline     
        In-The-Wild  &0.87 &0.79 &0.83 \\                  
       \hline
        ASVspoof5 Eva Subset &0.91 &0.91 &0.92\\     
        \hline             
        FakeAVCeleb &0.99 &0.99 &0.99\\                  
        \hline                         

    \end{tabular}
    }
    \vspace{-0.3cm}
    \label{table:res02} 
\end{table}

\subsection{Experimental Settings} 

To evaluate the backbone in the proposed DSD model, we evaluate three pre-trained models:  Whisper-base model~\cite{whisper}, WavLM-large~\cite{wavlm}, and Wave2XLSR~\cite{wav2vec20}, which show the state-of-the-art results on the upstream task of Speech-To-Text.
Regarding testing datasets, we evaluate different datasets of In-The-Wild~\cite{intwi_ds}, ASVspoof5 Evaluation Subset~\cite{asv15_best}, and FakeAVCeleb~\cite{FakeAVCeleb_s2}. The ASVspoof5 Evaluation~\cite{asv15_best} subset is selected as this dataset presents diverse bonafide resource and unseen fake generators compared with ASVspoof5 Development and Train subset in BR-AG dataset.
Meanwhile, FakeAVCeleb~\cite{FakeAVCeleb_s2} is selected as this dataset uses the VoxCeleb~\cite{voxceleb2} as the bonafide resource which presents real-world conversation from a large number of speakers. 
In this experiment, we report the Accuracy (Acc.), F1, and AuC scores using the the fixed probability of 0.5 as the boundary for all evaluating datasets. 
The other settings are re-used from the experiment in Section~\ref{first-exp}
\begin{figure}[t]
    \centering
    \scalebox{0.5}{
    \includegraphics[width=\textwidth]{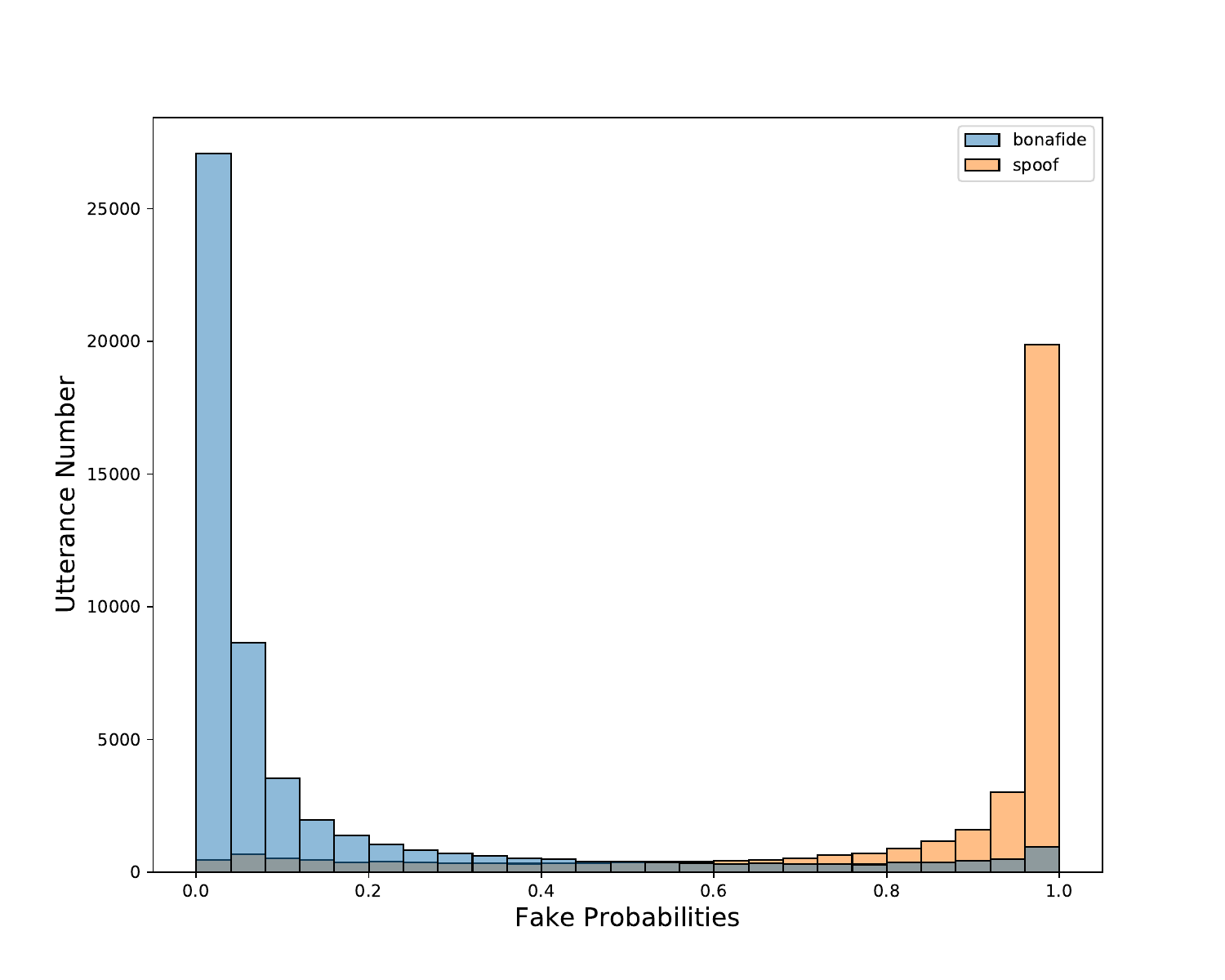}
    }
           \vspace{-0.7cm}
    \caption{Distribution of fake probabilities from In-The-Wild dataset using BR-AG dataset to train the WavLM-Large-Finetune+MLP model}
    \label{fig:dis05}    
\end{figure}

\begin{figure}[t]
    \centering
    \scalebox{0.5}{
    \includegraphics[width=\textwidth]{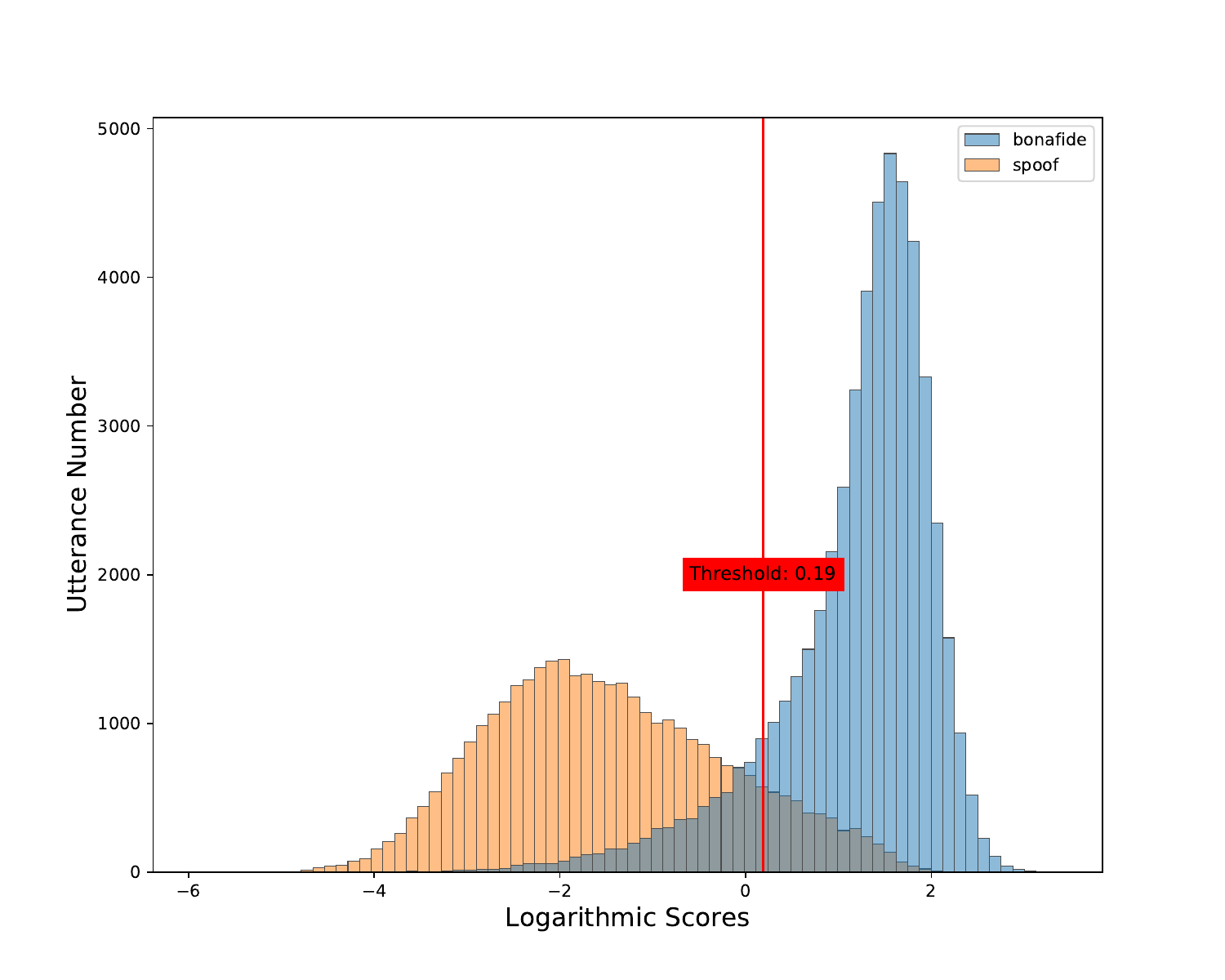}
    }
           \vspace{-0.7cm}
    \caption{Distribution of Logarithmic scores from In-The-Wild dataset using BR-AG dataset to train the WavLM-Large-Finetune+MLP model}
    \label{fig:dis06}    
\end{figure}

\subsection{Experimental Results and Discussion}

Compare among pre-trained models as shown in the lower part in the Table~\ref{table:res01}, this indicates that finetuning the pre-trained WavLM-Large with the three-stage training strategy is more effect than finetuning Wav2Vec2.0-XLSR-53 and Whisper-based models.
Finetuning the WavLM-Large backbone also enhances the performance compared with freezing the WavLM-large backbone as shown in the baseline.

Given the best model of WavLM-Large-Finetune+MLP, we evaluate this proposed model on different datasets.
As the experimental results are shown in the Table~\ref{table:res02}, the proposed model achieves accuracy scores of 0.87 and 0.91 on In-The-Wild and ASVspoof5 Evaluation Subset, respectively.
Significantly, the model achieves a high performance on FakeAVCeleb (i.e., All accuracy, F1, and AuC scores are  0.99). 
These results indicate that the proposed model is general on these three evaluating datasets as using the fixed boundary probability of 0.5.

The Fig.~\ref{fig:dis05} and Fig.~\ref{fig:dis06}, which present the distribution of fake probabilities and Logarithmic scores of In-The-Wild dataset using the best model of WavLM-Large-Finetune+MLP, also indicate that fake and bonafide distribution is well separated.

\section{Conclusion}
We have presented a comprehensive analysis of Bonafide Resource (BR) and AI-Based Generators (AG) and indicated how these two factors affect the performance and the generality of a Deefake Speech Detection (DSD) model.
Given the analysis, we successfully proposed the BR-AG dataset which is used to train and achieve a general DSD model.
Our proposed DSD model trained on the BR-AG dataset proves general on various benchmark DSD datasets and potential for real-work applications.
Our future work will explore cross-language evaluation.




\bibliographystyle{IEEEtran}
\bibliography{refs}

@article{Adam,
  title={Adam: A Method for Stochastic Optimization},
  author={Diederik,  P. K. and Jimmy,  B.},
  journal={CoRR},
  year={2015},
  volume={abs/1412.6980}
}

@article{wavlm,
  title={Wavlm: Large-scale self-supervised pre-training for full stack speech processing},
  author={Chen, Sanyuan and others},
  journal={IEEE Journal of Selected Topics in Signal Processing},
  volume={16},
  number={6},
  pages={1505--1518},
  year={2022}
}

@inproceedings{whisper,
  title={Robust speech recognition via large-scale weak supervision},
  author={Radford, Alec and others},
  booktitle={Proc. ICML},
  pages={28492--28518},
  year={2023}
}

@inproceedings{wav2vec20,
  author={Alexis Conneau and Alexei Baevski and Ronan Collobert and Abdelrahman Mohamed and Michael Auli},
  title={{Unsupervised Cross-Lingual Representation Learning for Speech Recognition}},
  year=2021,
  booktitle={Proc. INTERSPEECH},
  pages={2426--2430}
}

@inproceedings{muller2024mlaad,
  title={Mlaad: The multi-language audio anti-spoofing dataset},
  author={M{\"u}ller, Nicolas M and Kawa, Piotr and Choong, Wei Herng and Casanova, Edresson and G{\"o}lge, Eren and M{\"u}ller, Thorsten and Syga, Piotr and Sperl, Philip and B{\"o}ttinger, Konstantin},
  booktitle={Proc. IJCNN},
  pages={1--7},
  year={2024}
  }

@inproceedings{khalid2021fakeavceleb,
  title={FakeAVCeleb: A Novel Audio-Video Multimodal Deepfake Dataset},
  author={Khalid, Hasam and Tariq, Shahroz and Kim, Minha and Woo, Simon S},
  booktitle={Thirty-fifth Conference on Neural Information Processing Systems Datasets and Benchmarks Track (Round 2)},
  year={2021}
}

@misc{for_source,
  title={Audio Source Used to generate FoR Dataset},
  year={2018},
  howpublished={\url{https://www.kaggle.com/datasets/percevalw/englishfrench-translations}},
}

@article{dfdc_ds,
  title={The deepfake detection challenge ({DFDC}) dataset},
  author={Dolhansky, Brian and Bitton, Joanna and Pflaum, Ben and Lu, Jikuo and Howes, Russ and Wang, Menglin and Ferrer, Cristian Canton},
  journal={arXiv preprint arXiv:2006.07397},
  year={2020}
}

@inproceedings{VoC_ds,
  title={Spoofed training data for speech spoofing countermeasure can be efficiently created using neural vocoders},
  author={Wang, Xin and Yamagishi, Junichi},
  booktitle={Proc. ICASSP},
  pages={1--5},
  year={2023}
}

@inproceedings{intwi_ds,
  author={Nicolas Müller and Pavel Czempin and Franziska Diekmann and Adam Froghyar and Konstantin Böttinger},
  title={{Does Audio Deepfake Detection Generalize?}},
  year=2022,
  booktitle={Proc. INTERSPEECH},
  pages={2783--2787}
}

@article{wake_ds,
  title={Wavefake: A data set to facilitate audio deepfake detection},
  author={Frank, Joel and Sch{\"o}nherr, Lea},
  journal={NeurIPS},
  year={2024}
}

@article{cai2023av,
  title={AV-Deepfake1M: A Large-Scale LLM-Driven Audio-Visual Deepfake Dataset},
  author={Cai, Zhixi and Ghosh, Shreya and Adatia, Aman Pankaj and Hayat, Munawar and Dhall, Abhinav and Stefanov, Kalin},
  journal={arXiv preprint arXiv:2311.15308},
  year={2023}
}

@misc{1mdeepfake_ch,
  title={1M-Deepfakes Detection Challenge},
  year={2023},
  howpublished={\url{https://deepfakes1m.github.io/}},
}

@inproceedings{asv15_best,
  title={Combining evidences from mel cepstral, cochlear filter cepstral and instantaneous frequency features for detection of natural vs. spoofed speech.},
  author={Patel, Tanvina B and Patil, Hemant A},
  booktitle={Proc. INTERSPEECH},
  pages={2062--2066},
  year={2015}
}

@inproceedings{ASV15,
  title     = {ASVspoof 2015: the first automatic speaker verification spoofing and countermeasures challenge},
  author    = {Zhizheng Wu and Tomi Kinnunen and Nicholas Evans and Junichi Yamagishi and Cemal Hanilçi and Md. Sahidullah and Aleksandr Sizov},
  year      = {2015},
  booktitle = {Proc. INTERSPEECH},
  pages     = {2037--2041}
}

@article{ASV19,
  title={ASVspoof 2019: A large-scale public database of synthesized, converted and replayed speech},
  author={Wang, Xin and others},
  journal={Computer Speech \& Language},
  volume={64},
  pages={101114},
  year={2020}
  }

@inproceedings{ASV21,
  title={ASVspoof 2021: accelerating progress in spoofed and deepfake speech detection},
  author={Yamagishi, Junichi and others},
  booktitle={Workshop-Automatic Speaker Verification and Spoofing Coutermeasures Challenge (ASVspoof)},
  year={2021}
}

@misc{ASV24,
  title={The ASVspoof 2024 Challenge},
  year={2024},
  howpublished={\url{https://www.asvspoof.org/}},
}

@INPROCEEDINGS{for_ds,
  author={Reimao, Ricardo and Tzerpos, Vassilios},
  booktitle={International Conference on Speech Technology and Human-Computer Dialogue}, 
  title={FoR: A Dataset for Synthetic Speech Detection}, 
  year={2019},
  pages={1-10}
}

@inproceedings{f_ds_01,
  title={Ai-synthesized voice detection using neural vocoder artifacts},
  author={Sun, Chengzhe and Jia, Shan and Hou, Shuwei and Lyu, Siwei},
  booktitle={Proc. IEEE/CVF Conference on Computer Vision and Pattern Recognition},
  pages={904--912},
  year={2023}
}

@inproceedings{ljspeech_ds,
  title={Efficient neural audio synthesis},
  author={Kalchbrenner, Nal and Elsen, Erich and Simonyan, Karen and Noury, Seb and Casagrande, Norman and Lockhart, Edward and Stimberg, Florian and Oord, Aaron and Dieleman, Sander and Kavukcuoglu, Koray},
  booktitle={Proc. ICML},
  pages={2410--2419},
  year={2018}
}

@article{jsut_ds,
  title={JSUT corpus: free large-scale Japanese speech corpus for end-to-end speech synthesis},
  author={Sonobe, Ryosuke and Takamichi, Shinnosuke and Saruwatari, Hiroshi},
  journal={arXiv preprint arXiv:1711.00354},
  year={2017}
}

@inproceedings{voxceleb2,
  author={Joon Son Chung and Arsha Nagrani and Andrew Zisserman},
  title={{VoxCeleb2: Deep Speaker Recognition}},
  year=2018,
  booktitle={Proc. INTERSPEECH},
  pages={1086--1090}
}

@inproceedings{FakeAVCeleb_s2,
  title={A lip sync expert is all you need for speech to lip generation in the wild},
  author={Prajwal, KR and Mukhopadhyay, Rudrabha and Namboodiri, Vinay P and Jawahar, CV},
  booktitle={Proc. ACM international conference on multimedia},
  pages={484--492},
  year={2020}
}

@article{pham_01,
  title={A comprehensive survey with critical analysis for deepfake speech detection},
  author={Pham, Lam and Lam, Phat and Tran, Dat and Tang, Hieu and Nguyen, Tin and Schindler, Alexander and Skopik, Florian and Polonsky, Alexander and Vu, Hai Canh},
  journal={Computer Science Review},
  volume={57},
  pages={100757},
  year={2025}
  }

@inproceedings{libreispeech,
  title={Librispeech: an ASR corpus based on public domain audio books},
  author={Panayotov, Vassil and Chen, Guoguo and Povey, Daniel and Khudanpur, Sanjeev},
  booktitle={Proc. ICASSP},
  pages={5206--5210},
  year={2015}
  }

@article{novel_train,
  title={DIN-CTS: Low-complexity depthwise-inception neural network with contrastive training strategy for deepfake speech detection},
  author={Pham, Lam and Tran, Dat and Lam, Phat and Skopik, Florian and Schindler, Alexander and Poletti, Silvia and Fischinger, David and Boyer, Martin},
  journal={arXiv preprint arXiv:2502.20225},
  year={2025}
}

@article{sota_01_xlsr,
  title={XLSR-MamBo: Scaling the Hybrid Mamba-Attention Backbone for Audio Deepfake Detection},
  author={Ng, Kwok-Ho and Song, Tingting and WU, Yongdong and Xia, Zhihua},
  journal={arXiv preprint arXiv:2601.02944},
  year={2026}
}

@article{sota_02,
  title={WavLM model ensemble for audio deepfake detection},
  author={Combei, David and Stan, Adriana and Oneata, Dan and Cucu, Horia},
  journal={arXiv preprint arXiv:2408.07414},
  year={2024}
}

@article{sota_03,
  title={Speech df arena: A leaderboard for speech deepfake detection models},
  author={Dowerah, Sandipana and Kulkarni, Atharva and Kulkarni, Ajinkya and Tran, Hoan My and Kalda, Joonas and Fedorchenko, Artem and Fauve, Benoit and Lolive, Damien and Alum{\"a}e, Tanel and Doss, Matthew Magimai},
  journal={arXiv preprint arXiv:2509.02859},
  year={2025}
}
\end{document}